\documentclass[preprint,aps,prb]{revtex4}
\usepackage{epsfig}
\usepackage{float}

\begin{document}

\title{Transport properties in liquids from first principles: the case
of liquid water and liquid Argon}
\author{Pier Luigi Silvestrelli}

\affiliation{
Dipartimento di Fisica e Astronomia ``G. Galilei'',
Universit\`a di Padova, via Marzolo 8, I-35131 Padova, Italy}

\date{\today}

\begin{abstract}
Shear and bulk viscosity of liquid water and Argon are evaluated from first 
principles in the Density Functional Theory (DFT) framework, by performing
Molecular Dynamics simulations in the NVE ensemble and using
the Kubo-Greenwood equilibrium approach. 
Standard DFT functional is corrected in such a way to allow for a
reasonable description of van der Waals (vdW) effects.
For liquid Argon the thermal conductivity has been also calculated.
Concerning liquid water, to our knowledge this is the first 
estimate of the bulk viscosity and of the 
shear-viscosity/bulk-viscosity ratio from first principles.
By analyzing our results we can conclude that our first-principles simulations,
performed at a nominal average temperature of
366 K to guarantee that the systems is liquid-like, actually describe the basic 
dynamical properties of liquid water at about 330 K.
In comparison with liquid water, the normal, monatomic liquid Ar is 
characterized by a much smaller bulk-viscosity/shear-viscosity ratio 
(close to unity) and this feature is well reproduced by our 
first-principles approach which predicts a value of the ratio in 
better agreement with experimental reference data than
that obtained using the empirical Lennard-Jones potential.  
The computed thermal conductivity of liquid Argon is also in good agreement with
the experimental value. 

\end{abstract}
\maketitle

\vfill \eject

\section{Introduction}
Transport properties are among the most important and useful features
of condensed-matter systems, particularly for characterizing the dynamical
behavior of liquids, since they play an important role in many technical and 
natural processes. Therefore their estimate represents 
one of the most relevant goal of Molecular Dynamics (MD) simulation 
techniques which become particularly useful in cases where experimental data 
are not available or difficult to obtain. 
Different theoretical approaches can be adopted with a varying degree of 
accuracy (see, for instance, refs. \onlinecite{Allen,Helfand,Alder,Gosling,Ciccotti,Ciccotti78,Ciccotti79,Schoen,Hoheisel,Erpenbeck,Balasubramanian,Alfe,Guo,Viscardy,Jones,Lishchuk,Kim,Kirova,Shi,Chen,Rabani,Kusudo,Torres,Vogelsang,Fan,Kang,French,Fernandez,Marcolongo,Ercole,Marcolongo21,Tisi,Malosso,Ercole17,Ercole22}, and further references therein). 

Basically, in MD simulation transport properties can be evaluated 
either through a genuine {\it nonequilibrium approach} by applying
an explicit external perturbation (such as a shear flow or a temperature 
gradient), which is clearly
direct and intuitive but is affected by non-trivial technical issues
(in particular the need to generate nonequilibrium steady states in typical
systems characterized by finite-size supercells with periodic 
boundary conditions and to extrapolate to the limit of zero driving force).
Alternatively, the transport coefficients 
can be more easily estimated from {\it equilibrium} MD simulations by using 
the Green-Kubo relations\cite{Green,Kubo,McQuarrie} of statistical 
mechanics (dissipation-fluctuation theorem) which allow
the calculation of transport coefficients by integration of 
suitable autocorrelation functions. 
This latter approach is simpler because standard equilibrium MD simulations
can be easily carried out and estimated transport coefficients exhibit a weaker
system-size dependence.\cite{Kang}
An equivalent\cite{Kim} equilibrium method exploits the Einstein–Helfand 
expressions\cite{Helfand} to get transport coefficients directly from the 
particle displacements and velocities;\cite{Kirova} for instance, the
shear viscosity can be computed in terms of the mean-square
$x$ displacement of the center of $y$ momentum, while 
the thermal conductivity is proportional to the mean square $x$ displacement
of the center of energy.

The shear viscosity describes the resistance of a fluid to shear forces and 
is a measure of the shear stress induced by an
applied velocity gradient,\cite{Allen} while the bulk viscosity 
refers to the resistance to dilatation of an infinitesimal volume element 
at constant shape and measures the resistance of a fluid to compression.
It is closely connected with absorption and dispersion of ultrasonic waves in a 
fluid, so it can provide valuable information about intermolecular forces.
Moreover, the role of the bulk viscosity is acquiring more and more
importance, for instance in the area of surface and interface-related 
phenomena and for the interpretation of acoustic sensor data.\cite{Hafner} 
In spite of its relevance, bulk viscosity has received less experimental and
theoretical attention, partly due to the greater difficulties in obtaining
accurate measurements and estimates. 
In principle it should be evaluated in the microcanonical (NVE) ensemble 
where there is no need to evaluate an additional term
which would be required if, for instance, the canonical 
NVT ensemble were used.\cite{Guo,Hafner}
Moreover, bulk viscosity is subject to much larger statistical error
caused by the fact that it must be calculated by the
regression of fluctuations about a nonzero mean.\cite{Alder}
While the shear viscosity is associated with changes in water Hydrogen-bond
network connectivity and is mostly related to translational molecular motion, 
the bulk viscosity is associated with local density fluctuations and reflects the 
relaxation of both rotational and vibrational modes.\cite{Yahya,Dukhin}
The thermal conductivity describes instead the capability of a substance to 
allow molecular transport of energy driven by temperature gradients.

In general dynamical properties such as the transport coefficients are
much more dependent on the simulation size and timescale than
structural properties.\cite{Torres}
One must also point out that shear and bulk viscosities, 
and thermal conductivity 
are even more difficult to be evaluated accurately than, for instance, 
the diffusion coefficient (a single-particle property) since 
they are {\it collective} transport 
properties involving all the particles.\cite{Viscardy}
In fact, for estimating the diffusion coefficient one can perform
a statistical average over the particles in addition to the average
over time because every particle diffuses individually but any stress or 
energy fluctuation is an event involving the system as a whole. 
As a consequence, in order to obtain the same statistical 
accuracy, collective properties need much longer runs than single 
particle properties by a factor proportional to the size of the 
system.\cite{Alfe}

We here estimate from first principles simulations, in the framework of the 
Density Functional Theory (DFT), the shear and bulk viscosity of 
liquid water and Argon.
For liquid Argon the thermal conductivity is also calculated.
By analyzing our results we can conclude that our first-principles simulations,
performed at a nominal average temperature of
366 K to guarantee that the systems is liquid-like, actually describe the basic 
dynamical properties of liquid water at about 330 K.
Our approach is also able to reproduce well the bulk-viscosity/shear-viscosity ratio
of liquid Ar which is much smaller than that of liquid water.

\section{Method}
We have performed first principles MD simulations of liquid water using the
CPMD package,\cite{CPMD}
at constant volume, considering the experimental density of water at
room temperature.
The computations were performed at the $\Gamma$-point only of the
Brillouin zone, using norm-conserving pseudopotentials\cite{Troullier}
and a basis set of plane waves to expand the wavefunctions with an 
energy cutoff of 250 Ry; we have explicitly tested that this energy cutoff, 
much higher than that used in standard DFT simulations of liquid water, is required to 
have a good convergence also for the stress tensor components.

We have adopted the gradient-corrected BLYP\cite{BLYP}
density functional augmented by van der Waals (vdW) corrections,
hereafter referred to as
DFT-D2(BLYP).\cite{Grimme06}
This choice is motivated both by the fact that
BLYP has been shown\cite{Sprik96,PRL,JCP,Boero1,Boero2} to
give an acceptable description of Hydrogen bonding in water, and because
it represents a good reference DFT functional to add vdW 
corrections.\cite{PSIL1,PSIL2,PSIL3,Kannemann} 
A good description of Hydrogen bonding is essential here
since, in liquid water, the shear viscosity mostly originates from covalent
interactions in the Hydrogen-bond dynamics of water molecules.\cite{Shi}
Moreover, vdW corrections to BLYP are important because it was shown that
BLYP significantly underestimates (by 25\%) the equilibrium density of liquid water;
the experimental density can be recovered by adding the vdW corrections proposed
by Grimme,\cite{Grimme06} which have the further effect of making the oxygen-oxygen 
radial distribution function in better agreement with experiment.\cite{Schmidt,Wang}
Our system consists of 64 water molecules contained in a supercell
with simple-cubic symmetry and periodic boundary conditions.
Hydrogen nuclei have been treated as classical particles with the mass of the
deuterium isotope which allows us to use larger time steps.
The effective mass determining the time scale of the
fictitious dynamics of the electrons was 700 a.u. and the equations of motion
were integrated with a time step of 3 a.u. (=0.073 fs).


Our simulation consisted of an initial equilibration phase, lasting
about 0.15 ps, in which the ionic
temperature was simply controlled by velocity rescaling, followed by a 
much longer (about 22 ps) canonical (NVT) MD simulation 
(using suitable thermostats for a 
Nosé-Hoover dynamics), followed by a final 22 ps microcanonical (NVE) 
production MD run. 
A common drawback of most standard DFT functionals applied to liquid water at
room temperature is their tendency to ''freeze'' the system which therefore 
exhibits an ice-like behavior. By applying vdW corrections the problem
is reduced but it still present. In particular, since the melting temperature
of water estimated by DFT-D2(BLYP) is 360 K\cite{Yoo} (while it is 411 K with 
BLYP), following
a common strategy,\cite{Marcolongo,Tisi,Malosso,Grossman} we performed NVT simulations 
with an average ionic temperature
of 380 K to be sure that the system is indeed liquid-like. This use of
artificially increased temperature also serves to
mimic Nuclear Quantum Effects in simulations of liquid water.\cite{Torres}
The average ionic temperature of the subsequent NVE MD simulation was 366 K.
Several data (atomic coordinates, velocities, stress-tensor components,...) relevant 
for characterizing structural and dynamical properties of the system were recorded 
every 20 steps in the production stage.

As far as liquid Ar is concerned, before starting MD simulations,
we have performed extensive preliminary 
calculations to choose optimal parameters and a suitable DFT functional. 
Clearly in this case even an
empirical Lennard-Jones potential reference could probably give reasonable results
but here we are interested in studying transport properties using
DFT functionals in a first-principle framework, which has the advantage of
explicitly accounting for the electronic structure of matter.
Application to the face-centered cubic (fcc) Ar crystal (considering a fcc supercell with 32 
Ar atoms) and comparison with experimental reference values for the equilibrium
Ar-Ar distance and the cohesive energy, suggests that, among many tested,
vdW-corrected DFT functionals, DFT-D2(PBE)\cite{Grimme06,PBE} is the most adequate to describe
extended systems made by Ar atoms, hence we mainly use it for the MD 
simulations of liquid Ar. In this case we have checked that a suitable 
energy cutoff to get a good convergence for the stress tensor components is
110 Ry.

The liquid Ar sample was prepared starting from an initial (unfavorable) 
simple cubic lattice configuration with 64 Ar atoms and considering 
the experimental Ar density (1.4 g/cm$^3$) at melting point (84 K). 
Then the systems was heated by gradually increasing the ionic temperature 
(by velocity rescaling) to 500 K (in a time of 1.3 ps) to be sure that 
the system was truly melted; then the temperature was gradually decreased
(in 1.0 ps) to 150 K, which is a temperature sufficiently higher than the
experimental melting point that it can be assumed that the system is indeed 
in a liquid phase; this has been explicitly checked looking at the translational 
order parameter.\cite{Allen}

Then a 60 ps canonical (NVT) MD simulation (with a ionic temperature of 150 K)
was performed, followed by a 60 ps microcanonical (NVE) MD production runs with an average ionic temperature of 129 K.
In this case the electronic effective mass was 700 a.u. and the equations 
of motion were integrated with a time step of 5 a.u. (=0.121 fs).
Data (atomic coordinates, velocities, stress-tensor components,...) relevant 
for structural and dynamical properties of the system were recorded every 
10 steps in the production stage.

As mentioned above, different approaches exist for the calculation of shear, 
$\eta_S$, and bulk, $\eta_B$, viscosity from MD 
simulations.\cite{Allen,Guo,Balasubramanian,Erpenbeck,Schoen,Hoheisel}
The most used technique is based on the evaluation of the autocorrelation 
functions of stress-tensor components; in particular,\cite{Allen} 

\begin{equation}
\eta_S = {V\over {k_B T}} \int_0^{\infty} dt 
\langle P_{\alpha\beta}(0)P_{\alpha\beta}(t)\rangle \;,
\label{shearvisc}
\end{equation} 

\begin{equation}
\eta_B = {V\over {9k_B T}} \int_{0}^{\infty} dt 
\langle \delta P_{\alpha\alpha}(0) \delta P_{\beta\beta}(t)\rangle = 
{{V}\over {k_B T}} \int_{0}^{\infty} dt
\langle \delta P(0) \delta P(t)\rangle \;,
\label{bulkvisc}
\end{equation} 


where, in practice the upper limit of integration ($\infty$) is 
replaced by a reasonably-long simulation time, $t_{max}$, 
$\langle ... \rangle$ denotes average over different time origins,
$V$ is the system volume, $T$ the ionic temperature, $k_B$ the Boltzmann
constant, $P_{\alpha\beta}$ quantities denote the components of the stress tensor,
the instantaneous pressure is given by $P(t)=1/3 \sum_{\alpha} P_{\alpha\alpha}$
(that is the average of the diagonal elements of the stress tensor), and the
fluctuations are defined as:

\begin{equation}
\delta P_{\alpha\alpha}(t) = P_{\alpha\alpha}(t) - \langle P_{\alpha\alpha} \rangle
= P_{\alpha\alpha}(t) - P \;,
\delta P(t) = P(t) - \langle P \rangle = P(t) - P \;,
\label{fluctuations}
\end{equation}

where $P$ is the system pressure obtained as the ensemble average of $P(t)$.
In isotropic fluids (with rotational invariance) there are only 5 independent
(and equivalent) components of the traceless stress tensor:
$P_{xy}$, $P_{yz}$, $P_{zx}$, $(P_{xx}-P_{yy})/2$, and $(P_{yy}-P_{zz})/2$,
so that it is convenient to compute the shear viscosity $\eta_S$ by averaging
over these 5 components to get better statistics.

Instead, the bulk viscosity $\eta_B$ has only one component, moreover
the diagonal stresses must be evaluated carefully since a 
non-vanishing equilibrium average must be subtracted.
In oder to get more accurate evaluations of transport properties and
also reliable estimates of the associated statistical errors,
we adopt the block-average technique,\cite{Frenkel} which consists
of dividing the whole simulation into a sequence of several shorter
intervals (``blocks''), each with an equal number
of samples; then block averages are calculated which allow to 
estimate means and variances.\cite{Jones}
In the case of the bulk-viscosity calculation,
to reduce the error, it is convenient to take for the system pressure
the average value of the pressures over all blocks.\cite{Guo}    
Clearly the choice of the block size must be made with care; in fact,
samples become uncorrelated as the block size increases so
for small block sizes, the error is underestimated while
for large block sizes the error estimate is inaccurate due to
insufficient sampling (see detailed discussion below).

Typically transport coefficients are estimated from classical MD 
simulations based on empirical interatomic potentials.
The practical feasibility of calculating transport coefficients in liquids 
using instead first principles MD simulations, was demonstrated
by D. Alf\'e and M. J. Gillan,\cite{Alfe} who used the Green-Kubo 
relations to compute the shear 
viscosity of liquid iron and aluminum, with a statistical error of
about 5\%. However, the simulations of D. Alf\'e and M. J. Gillan\cite{Alfe} 
were performed in the NVT ensemble, while our simulations have been
carried out using the NVE ensemble, since the NVE simulations
also allow the evaluation of the bulk viscosity without any 
correction term (see above).
Subsequently, the problem of atomic heat transport and the development of a 
microscopic theory for the calculation of the thermal
conductivity within the Green–Kubo formalism, by a first-principles DFT approach, has 
been successfully addressed by Marcolongo {\it et al.}\cite{Marcolongo,Ercole,Marcolongo21} 
Efficient, first-principles evaluations of thermal conductivity and shear viscosity
have been also recently proposed.\cite{Tisi,Malosso,Ercole17,Ercole22}

A simpler alternative method exists (valid for temperatures that are not 
too low\cite{Herrero}) to obtain an approximate estimate of 
the shear viscosity, by exploiting its connection with the self-diffusion 
coefficient $D$ via the Stokes-Einstein relation:\cite{Gosling,Alfe}

\begin{equation}
\eta_S = {{k_B T}\over {2\pi a D}}\;,  
\label{stokeseins}
\end{equation} 

where $a$ is an effective atomic diameter. Such relation is exact
for the Brownian motion of a macroscopic particle of
diameter $a$ in a liquid of shear viscosity $\eta_S$, but it is only approximate when
applied to atoms; however if $a$ is chosen to be the radius $r_1$ of the first
peak in the radial distribution function, the relation usually predicts $\eta_S$
to within 40\%.\cite{Alfe}
Here we take for $r_1$ the position of the first
peak in the O-O and Ar-Ar radial distribution function for liquid water and
liquid Ar, respectively, while  
the diffusion coefficient $D$ can be computed\cite{Allen} 
from the mean square displacement of the oxygen atoms (for liquid water) or
Ar atoms (for liquid Ar).
The validity of the Stokes–Einstein relation has been recently discussed in detail by
Herrero {\it et al.}\cite{Herrero} who also explored the connection
between structural properties and transport coefficients.

For liquid Argon the thermal conductivity has been also calculated,
using the formula:\cite{Allen}

\begin{equation}
\lambda_T = {V\over {k_B T^2}} \int_{0}^{\infty} dt 
\langle j_{\alpha}^E(0) j_{\alpha}^E(t) \rangle \;,
\label{thermcond}
\end{equation} 

where $j_{\alpha}^E$ is the $\alpha$ component of the energy current defined as
the time derivative of 

\begin{equation}
\delta E_{\alpha} = {1\over V} \sum_i r_{i\alpha}(E_i - \langle E_i \rangle)\;,
\label{deltae}
\end{equation}

and $E_i$ is the energy of the $i-th$ Ar atom (located at coordinates 
$r_{ix}$, $r_{iy}$, $r_{iz}$), which can be evaluated as 

\begin{equation}
E_i = p_i^2/2m_i + 1/2 \sum_{j\neq i} v(r_{ij})\;,
\label{Ei}
\end{equation}

by assuming a {\it pairwise} interatomic potential. 
In order to obtain a pair potential
for evaluating the thermal conductivity of liquid Ar using configurational data
from our first-principles DFT simulations, we have adopted a strategy 
similar to that proposed in ref. \onlinecite{Kang}: 
we assume for the pair potential a Lennard-Jones analytical form:

\begin{equation}
v(r) = a (b^2/r^{12} - b/r^6)\;,
\label{ljpot}
\end{equation}

where $a$ and $b$ are parameters optimized by fitting the potential-energy curve of
the Ar dimer (at different interatomic distances) obtained by using our DFT approach.  
Admittedly, such an approach is rather approximated and could represent only a
rough estimate of the thermal conductivity. A more fundamental and accurate 
method has been recently proposed.\cite{Marcolongo,Ercole,Marcolongo21} 

\section{Results and Discussion}
In Fig. 1 and 2 we plot the behavior of the temperature and pressure as a function
of time in the NVE simulation for liquid water and Ar, respectively. 
As can be seen, these quantities turn out to be stable
and exhibit only moderate oscillations around the average values, which are,
for liquid water,
0.132 GPa and 366 K for the pressure and the temperature, respectively, while
for liquid Ar the values are 0.173 GPa and 129 K.

\pagestyle{empty}
\begin{figure}
\centerline{
\includegraphics[width=17cm,angle=270]{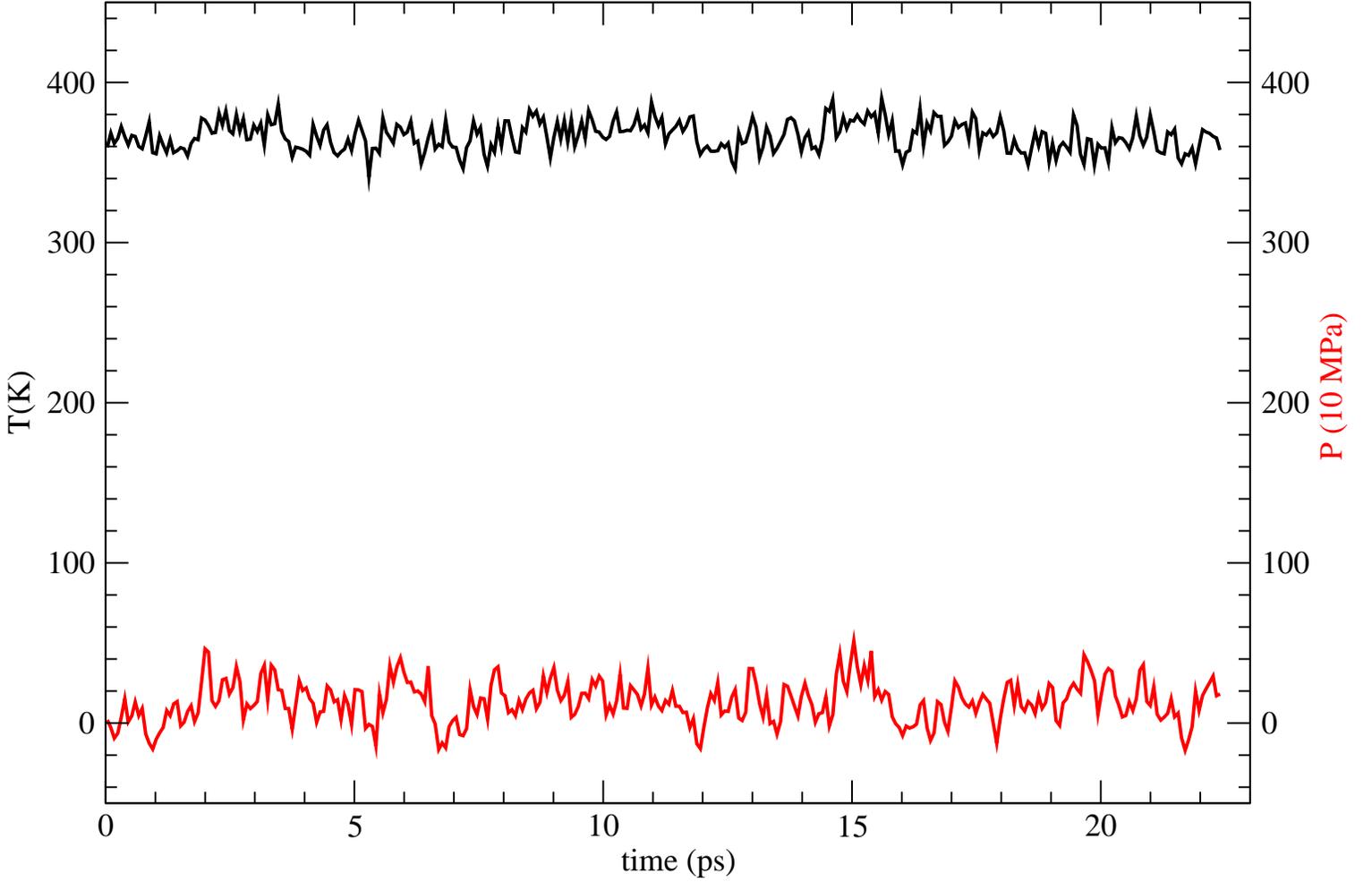}
}
\caption{Temperature and pressure of liquid water plotted as a function of 
the simulation time.}
\label{fig1}
\huge
\end{figure}

\pagestyle{empty}
\begin{figure}
\centerline{
\includegraphics[width=17cm,angle=270]{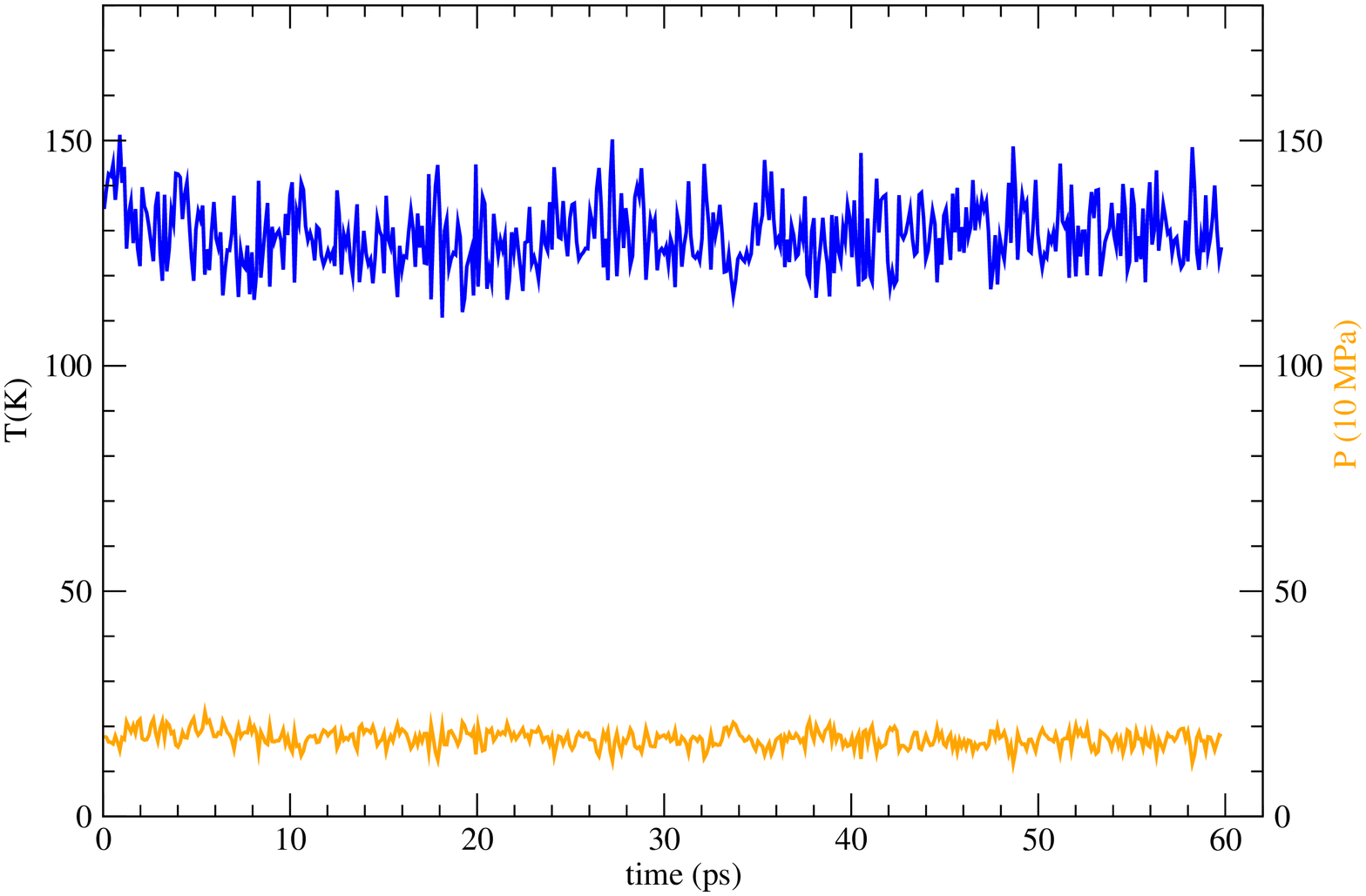}
}
\caption{Temperature and pressure of liquid Ar plotted as a function of 
the simulation time.}
\label{fig2}
\huge
\end{figure}

\pagestyle{empty}
\begin{figure}
\centerline{
\includegraphics[width=17cm,angle=270]{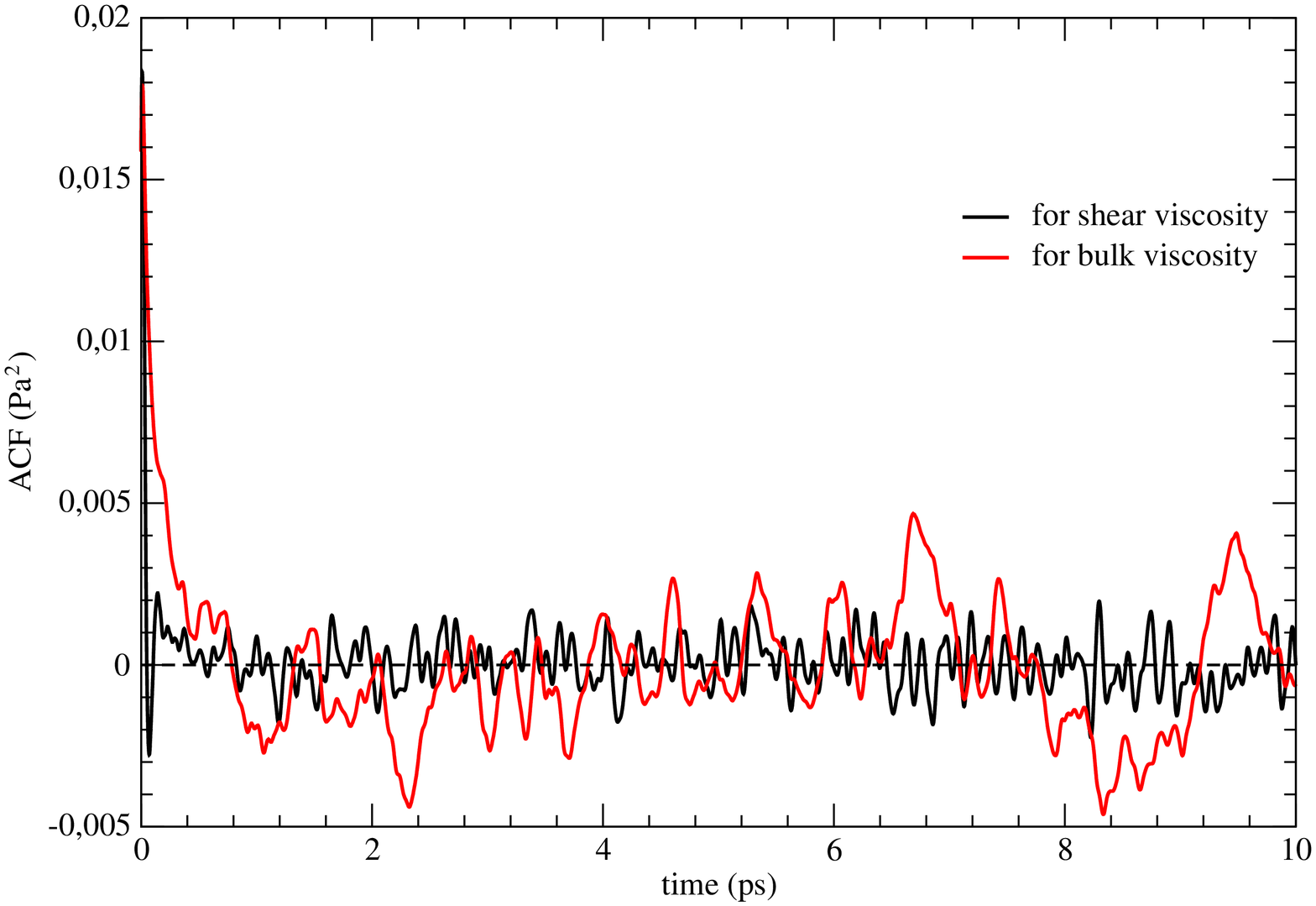}
}
\caption{Auto-correlation functions (ACFs) used for the evaluation of the shear and
bulk viscosities of liquid water (see text) plotted as a function of 
the simulation time.}
\label{fig3}
\huge
\end{figure}

\pagestyle{empty}
\begin{figure}
\centerline{
\includegraphics[width=17cm,angle=270]{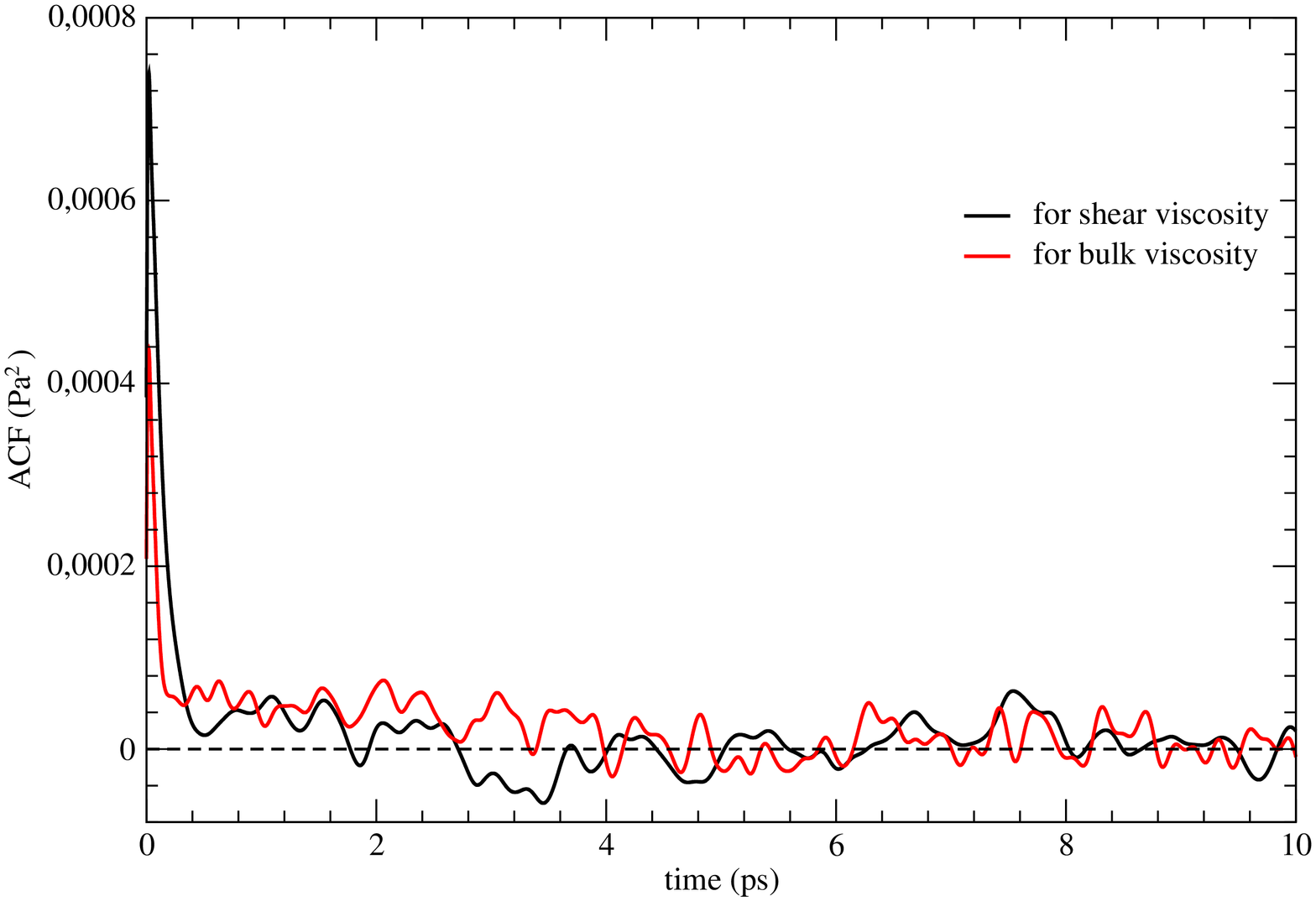}
}
\caption{Auto-correlation functions (ACFs) used for the evaluation of the shear and
bulk viscosities of liquid Ar (see text) plotted as a function of 
the simulation time.}
\label{fig4}
\huge
\end{figure}

In Fig. 3 and 4 we instead plot the auto-correlation functions (ACFs),
corresponding to the integrands (considering the average over the components 
for the shear viscosity) of eqs. (\ref{shearvisc}) and (\ref{bulkvisc}).
Differently from what observed in monatomic systems (such as liquid Ar)
or in classical MD simulations where waters are modeled by rigid molecules,
in first-principles simulations of liquid water, high-frequency 
intermolecular vibrations lead to corresponding high-frequency oscillations in the 
pressure and in related ACFs. In order to better appreciate the global
decay behavior of ACFs, in the case of liquid water,
high-frequency components have been cut by Fourier-transforming the
ACFs. A quantitative estimate of the ACFs relaxation times can be
obtained assuming a global exponential decay ($\simeq e^{-t/\tau}$)
of the integrands and computing:
\begin{equation}
\tau_S = \int_0^{\infty} dt 
{{\langle P_{\alpha\beta}(0)P_{\alpha\beta}(t)\rangle}
\over {\langle P_{\alpha\beta}(0)P_{\alpha\beta}(0)\rangle}} \;,
\label{taushearvisc}
\end{equation} 
and
\begin{equation}
\tau_B = \int_{0}^{\infty} dt 
{{\langle \delta P_{\alpha\alpha}(0) \delta P_{\beta\beta}(t)\rangle}
\over {\langle \delta P_{\alpha\alpha}(0) \delta P_{\beta\beta}(0)\rangle}} 
\label{taubulkvisc}
\end{equation} 

For liquid water we find $\tau_S \simeq 6$ fs and $\tau_B \simeq 4$ fs, while for liquid
Ar $\tau_S \simeq 340$ fs and $\tau_B \simeq 410$ fs.  

\pagestyle{empty}
\begin{figure}
\centerline{
\includegraphics[width=17cm,angle=270]{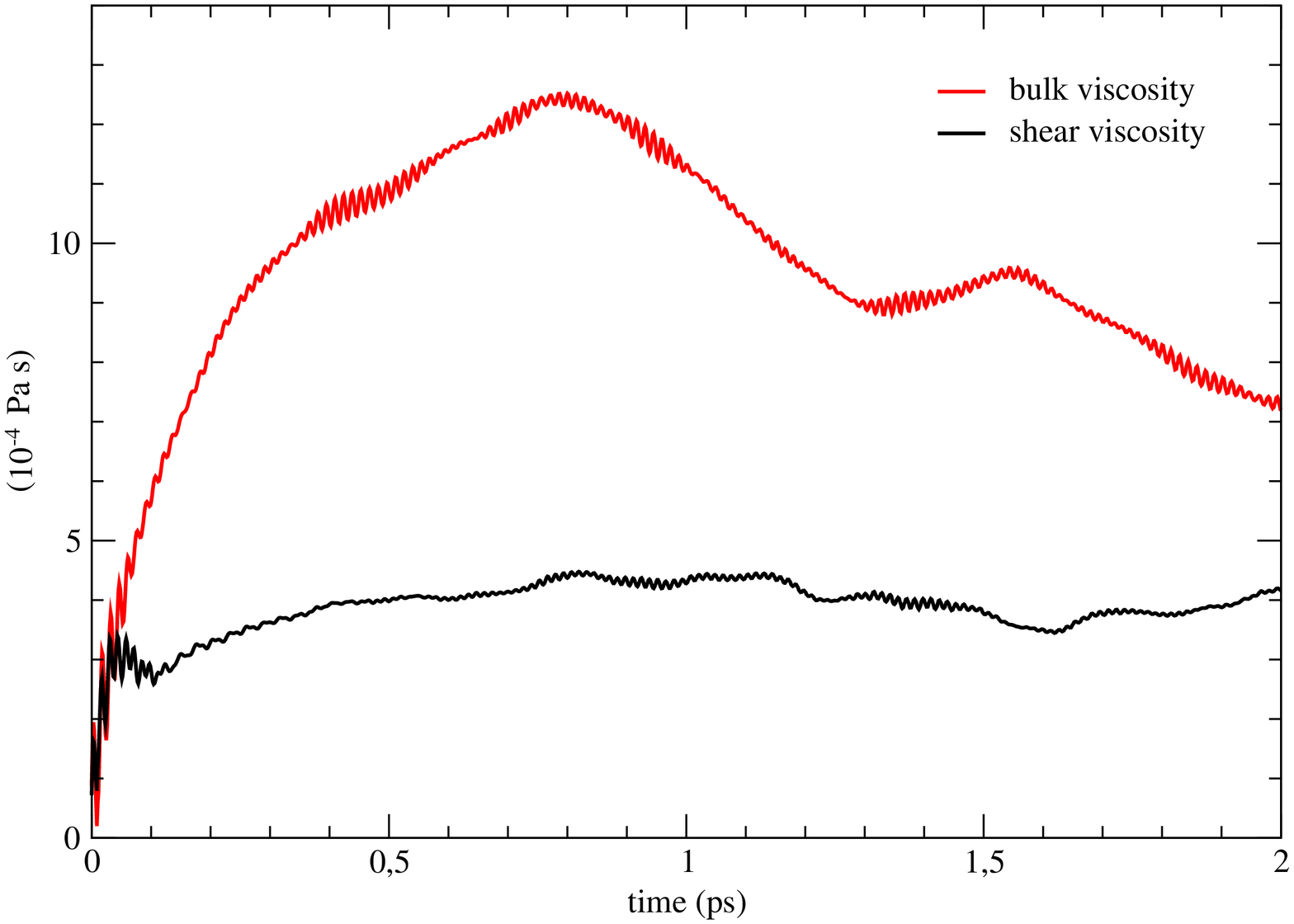}
}
\caption{Shear and bulk viscosity of liquid water plotted as a function of 
the upper limit of the integrals of the ACFs.}
\label{fig5}
\huge
\end{figure}

\pagestyle{empty}
\begin{figure}
\centerline{
\includegraphics[width=17cm,angle=270]{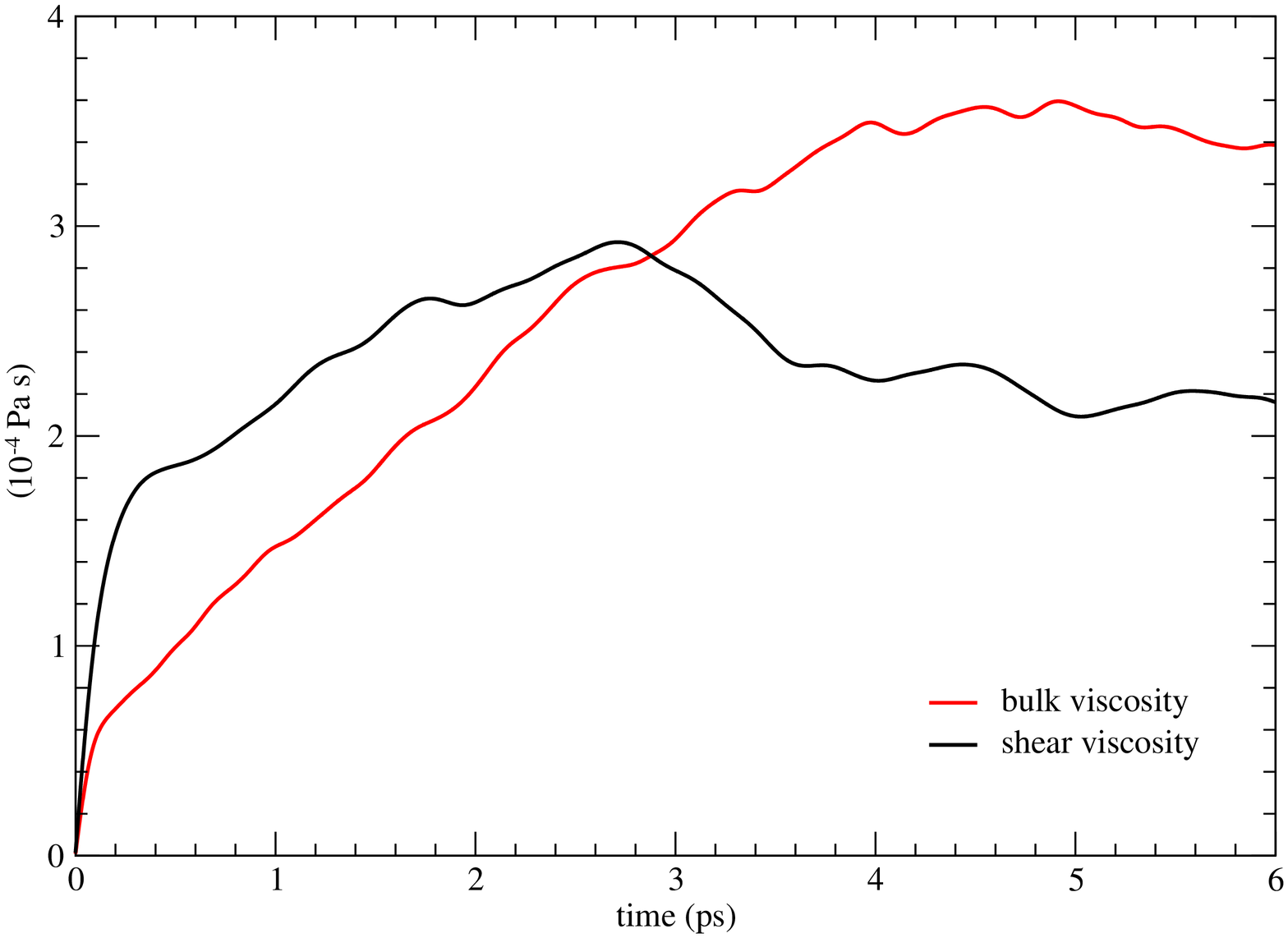}
}
\caption{Shear and bulk viscosity of liquid Ar plotted as a function of 
the upper limit of the integrals of the ACFs.}
\label{fig6}
\huge
\end{figure}

\pagestyle{empty}
\begin{figure}
\centerline{
\includegraphics[width=17cm,angle=270]{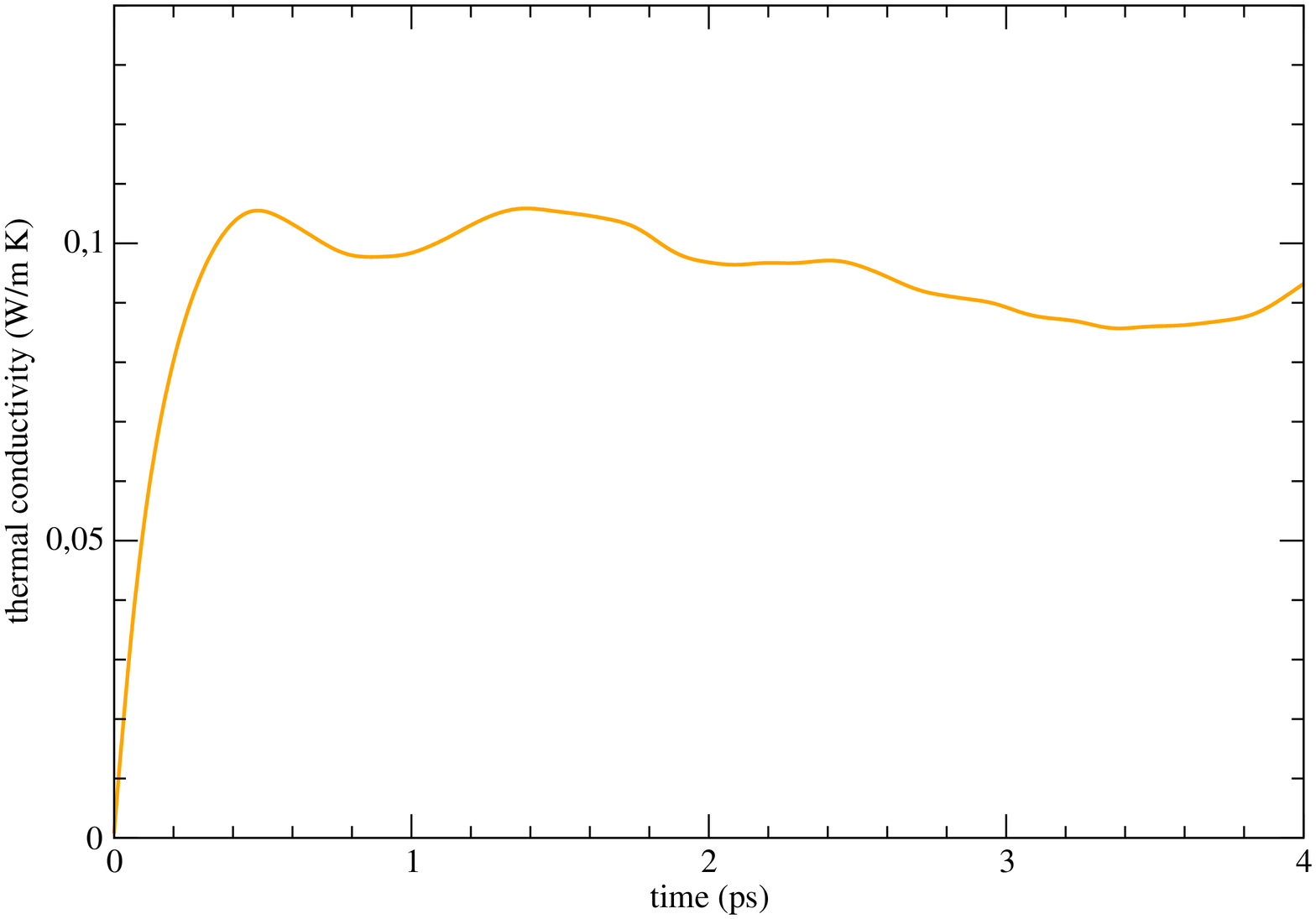}
}
\caption{Thermal conductivity of liquid Ar plotted as a function of 
the upper limit of the integral of the ACF.}
\label{fig7}
\huge
\end{figure}

The shear and bulk viscosity, computed using eqs. (\ref{shearvisc}) and 
(\ref{bulkvisc}), are plotted as a function of the upper limit
of the integrals in Fig. 5 and 6, while the thermal conductivity 
of liquid Ar is reported in Fig. 7.  
From these curves an approximate estimate of the shear and bulk viscosity
can be obtained considering the values of the quantities corresponding
to the position of the first pronounced maximum-plateau; in fact this
indicates that the running integral starts becoming nearly independent of time
implying that the corresponding ACF has decayed to zero and is fluctuating 
along the horizontal time axis. Clearly, considering longer times only
introduces additional noise to the signal and the beginning of a plateau
represents the desired value of the viscosity with the smallest uncertainty.
As can be seen, the maximum-plateau is reached at about $t=0.8$ ps
for both the shear and bulk viscosity of liquid water, while
the corresponding values for liquid Ar are 3.0, 5.0 ps, and 0.5 ps for the
shear viscosity, the bulk viscosity, and the thermal conductivity, respectively. 
As expected, these times are much larger than the corresponding relaxation times 
$\tau_S$ and $\tau_B$ estimated above. 

As already discussed, a more accurate evaluation, with also a reliable estimate
of the associated statistical error, can be obtained by adopting a
block-average technique. In this case a proper choice of the block size is
crucial: with many, small-size blocks, the statistical error is small
but the blocks are probably correlated and the viscosity is typically
underestimated (not yet converged); on the contrary, with just a few, 
large-size blocks, these are probably uncorrelated and
the viscosity is converged but the statistical error is large.   

In Figs. 8, 9, 10, and 11 we plot the values of the shear and bulk viscosity 
of liquid water and Ar evaluated by using different numbers of blocks 
(keeping constant the total number
of configurations) with the relative statistical errors.
The dashed horizontal lines indicate the corresponding values inferred by 
considering the maximas-plateaus of the curves in Figs. 5 and 6.
As can be seen, in the case of liquid water, the maxima of the shear and 
bulk viscosities are obtained 
considering 16 blocks, each equivalent to a simulation time of about 1.4 ps.
Interestingly, taking statistical uncertainties into account, these maxima
are compatible with the rough estimates obtained before and, for the
shear viscosity, also with the values obtained using the Stokes-Einstein 
formula (Eq. (\ref{stokeseins})). 
As already described above, 
in the Stokes-Einstein estimate the shear viscosity is obtained in terms of the
diffusion coefficient $D$ and the radius of the first
peak in the radial distribution function (see Eq. (\ref{stokeseins}), for liquid water
we have considered the first peak in the oxygen-oxygen radial distribution function, see
below). 
Actually our reported Stokes-Einstein estimated values are corrected by 
finite-size effects: in fact $D$ can be extrapolated to infinite size of the
simulation box (see, for instance, ref. \onlinecite{Herrero}) by just considering the
shear-viscosity value:

\begin{equation}
D_{\infty} = D + 2.837 {{k_B T}\over {6\pi \eta_S L}}\;,
\label{dcorrected}
\end{equation} 

where $L$ is the size of the cubic simulation box.
Therefore, by simultaneously taking into account Eqs. (\ref{stokeseins}) and 
(\ref{dcorrected}), one can get a ``self-consistent'', finite-size
corrected Stokes-Einstein estimate for $\eta_S$ : 

\begin{equation}
\eta_S^* = {{k_B T}\over {2\pi a D}} - 2.837 {{k_B T}\over {6\pi L D}}\;.  
\label{stokeseinscorrected}
\end{equation} 

\pagestyle{empty}
\begin{figure}
\centerline{
\includegraphics[width=17cm,angle=270]{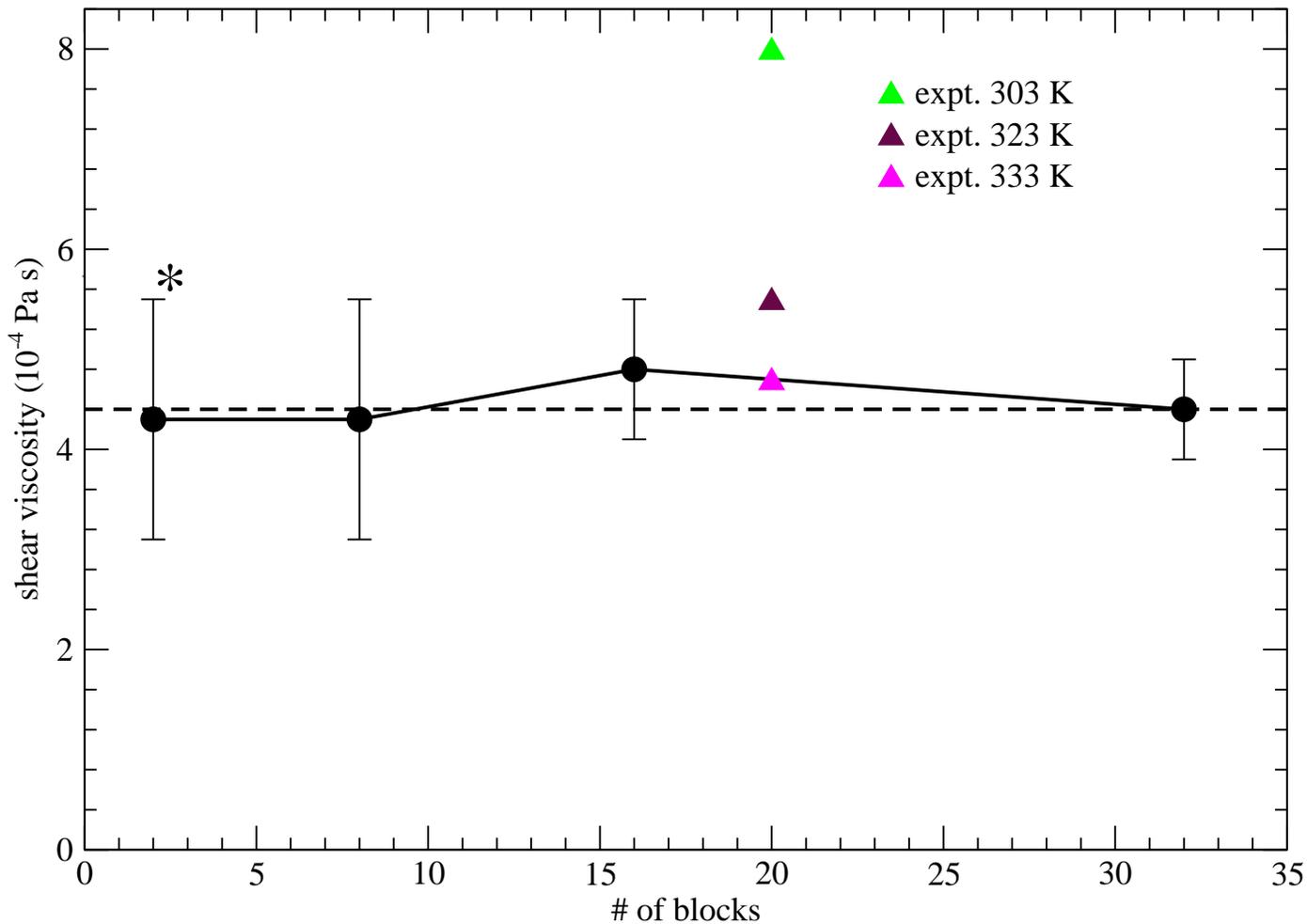}
}
\caption{Shear viscosity of liquid water evaluated by using 
different numbers of blocks (the smaller is the block number the larger 
is the number of configurations of each block) with the relative 
statistical errors.
The dashed horizontal line indicates the position of the first-pronounced 
maximum-plateau of the corresponding curve of Fig. 5. The asterisk denotes the value
obtained by the Stokes-Einstein formula (Eq.\ref{stokeseinscorrected}), while the
triangles indicate experimental estimates at different temperatures.}
\label{fig8}
\huge
\end{figure}

\begin{figure}
\centerline{
\includegraphics[width=17cm,angle=270]{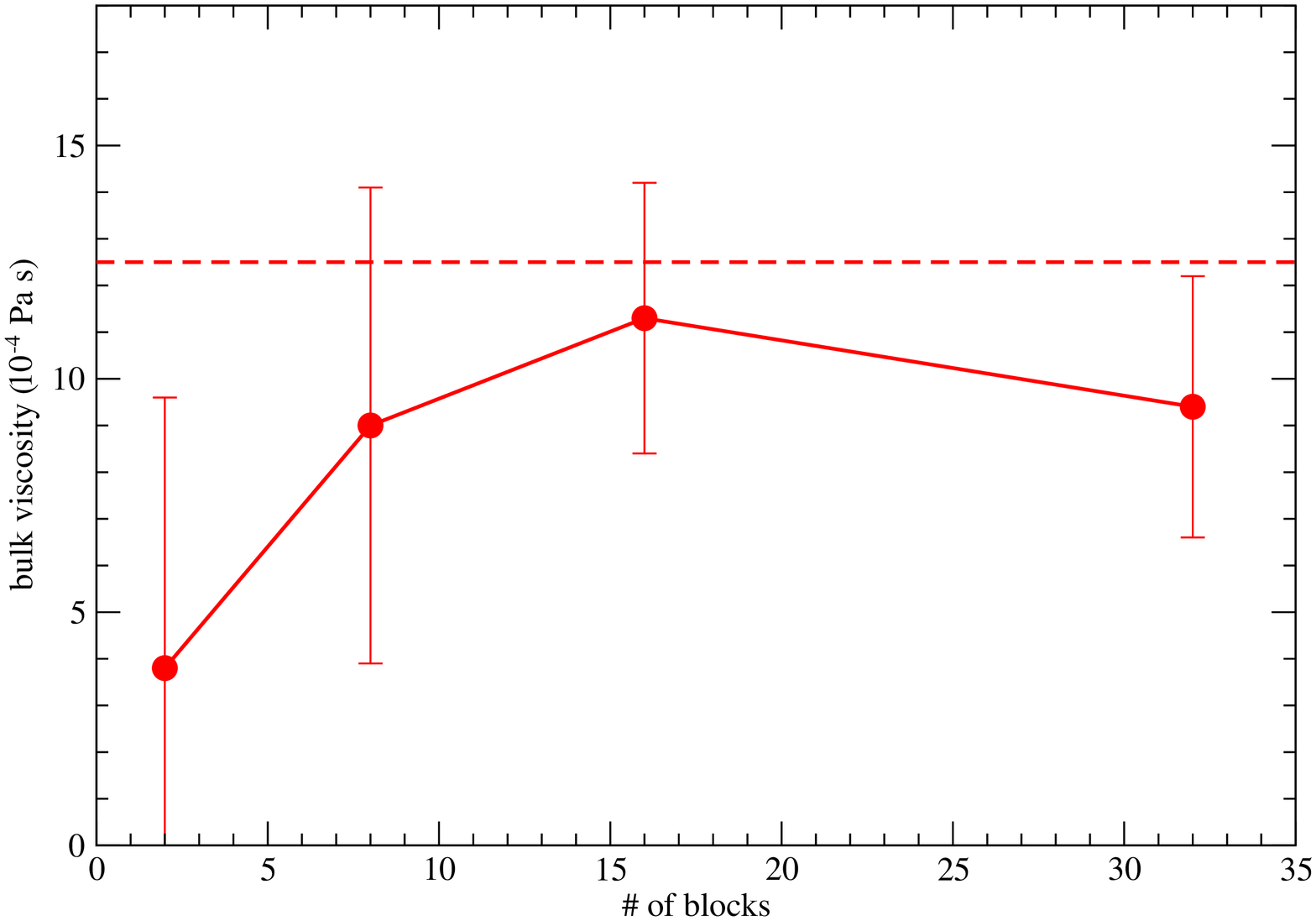}
}
\caption{Bulk viscosity of liquid water evaluated by using 
different numbers of blocks (the smaller is the block number the larger 
is the number of configurations of each block) with the relative 
statistical errors.
The dashed horizontal line indicates the position of the first-pronounced
maximum-plateau of the corresponding curve of Fig. 5.} 
\label{fig9}
\huge
\end{figure}

Quantitative data are collected in Table I where they are also compared with some theoretical
and experimental reference values.

\begin{table}
\vfill
\eject
\caption{Shear and bulk viscosity of liquid water and Ar, in $10^{-4}$ Pa s, 
compared with theoretical and experimental reference data. 
Statistical errors are in parenthesis. $\eta_S^*$ indicates the
shear viscosity estimate obtained by the Stokes-Einstein relation (see text).}
\begin{center}
\begin{tabular}{|c|c|c|c|c|c|}
\hline
system & $\eta_S$ & $\eta_S^*$& $\eta_B$ & $\eta_B/\eta_S$ & $3/4\eta_B/\eta_S +1$ \\ \tableline
\hline
 water (366 K) & 4.8(0.7) & 5.7 &11.3(2.9) & 2.4(0.8) & 2.8(0.6) \\
\hline
  water DFT SCAN$^a$ (300K)       &23    & --- & ---      & ---      & ---      \\
  water DFT SCAN$^a$ (330K)       & 6    & --- & ---      & ---      & ---      \\
  water DFT SCAN$^a$ (360K)       & 5    & --- & ---      & ---      & ---      \\
  water DFT OPTB88-vdW$^a$ (300K) &30    & --- & ---      & ---      & ---      \\
  water DFT OPTB88-vdW$^a$ (330K) &15    & --- & ---      & ---      & ---      \\
  water DFT OPTB88-vdW$^a$ (360K) & 8    & --- & ---      & ---      & ---      \\
  water force field$^a$ (300K)    & 8    & --- & ---      & ---      & ---      \\
  water force field$^a$ (330K)    & 5    & --- & ---      & ---      & ---      \\
  water force field$^a$ (360K)    & 3.5  & --- & ---      & ---      & ---      \\
  water force field$^b$ (303K)    & 6.5(0.4)&---&15.5(1.6)& 2.4(0.3) & 2.8(0.2) \\
\hline
 water expt.$^c$ (298 K)          & 8.90     & --- & ---      & ---      & ---      \\ 
 water expt.$^{b,d,e}$ (303 K)    & 7.97     & --- & 21.5     & 2.7      & 3.0      \\ 
 water expt.$^f$ (323 K)          & 5.47     & --- & 14.8     & 2.7      & 3.0      \\ 
 water expt.$^c$ (333 K)          & 4.67     & --- & ---      & ---      & ---      \\ 
\hline
 Ar (129 K)              & 3.7(1.6) & 2.0 & 4.0(2.2) & 1.1(0.8) & 1.8(0.6) \\
 Ar expt.$^g$ (90 K)     & 2.33     & --- & 1.82     & 0.8      & 1.6      \\
 Ar expt.$^h$ (90 K)     & 2.57     & --- & ---      & ---      &          \\
\hline
\end{tabular}                                                
\tablenotetext[1]{ref.\onlinecite{Herrero}.}
\tablenotetext[2]{ref.\onlinecite{Guo}.}
\tablenotetext[3]{ref.\onlinecite{Harris}.}
\tablenotetext[4]{ref.\onlinecite{Litovitz}.}
\tablenotetext[5]{ref.\onlinecite{Sengers}.}
\tablenotetext[6]{ref.\onlinecite{Holmes}.}
\tablenotetext[7]{ref.\onlinecite{Cowan}.}
\tablenotetext[8]{ref.\onlinecite{NIST}.}
\end{center}
\label{table1}                                  
\end{table}

As far as the shear viscosity is concerned, for liquid water our estimated
value, obtained from the NVE simulation at an average temperature of 366 K,
agrees with the experimental reference data at a lower temperature
of about 330 K. This is in line
with the performances of other DFT functionals; for instance (see Table I), in 
recent simulations\cite{Herrero} of liquid water based on the SCAN functional,\cite{SCAN}
the shear viscosity estimate is close to that obtained from a force-field
approach (that, for this quantity, well reproduces the experimental behavior)
only between 330 and 360 K, while it is severely overestimated at 300 K.   
A similar behavior has been found for the SCAN functional by 
Malosso {\it et al.},\cite{Malosso} who computed the shear viscosity of liquid water
from first-principles and deep-neural-network simulations.
With the OPTB88-vdW functional\cite{OPTB88-vdW} reasonable agreement with experimental
data at room temperature is only found\cite{Herrero} at 360 K, while with the
PBE functional\cite{PBE} this is only true at 454 K.\cite{Malosso}

\pagestyle{empty}
\begin{figure}
{\vskip -3.0cm}
\centerline{
\includegraphics[width=17cm,angle=270]{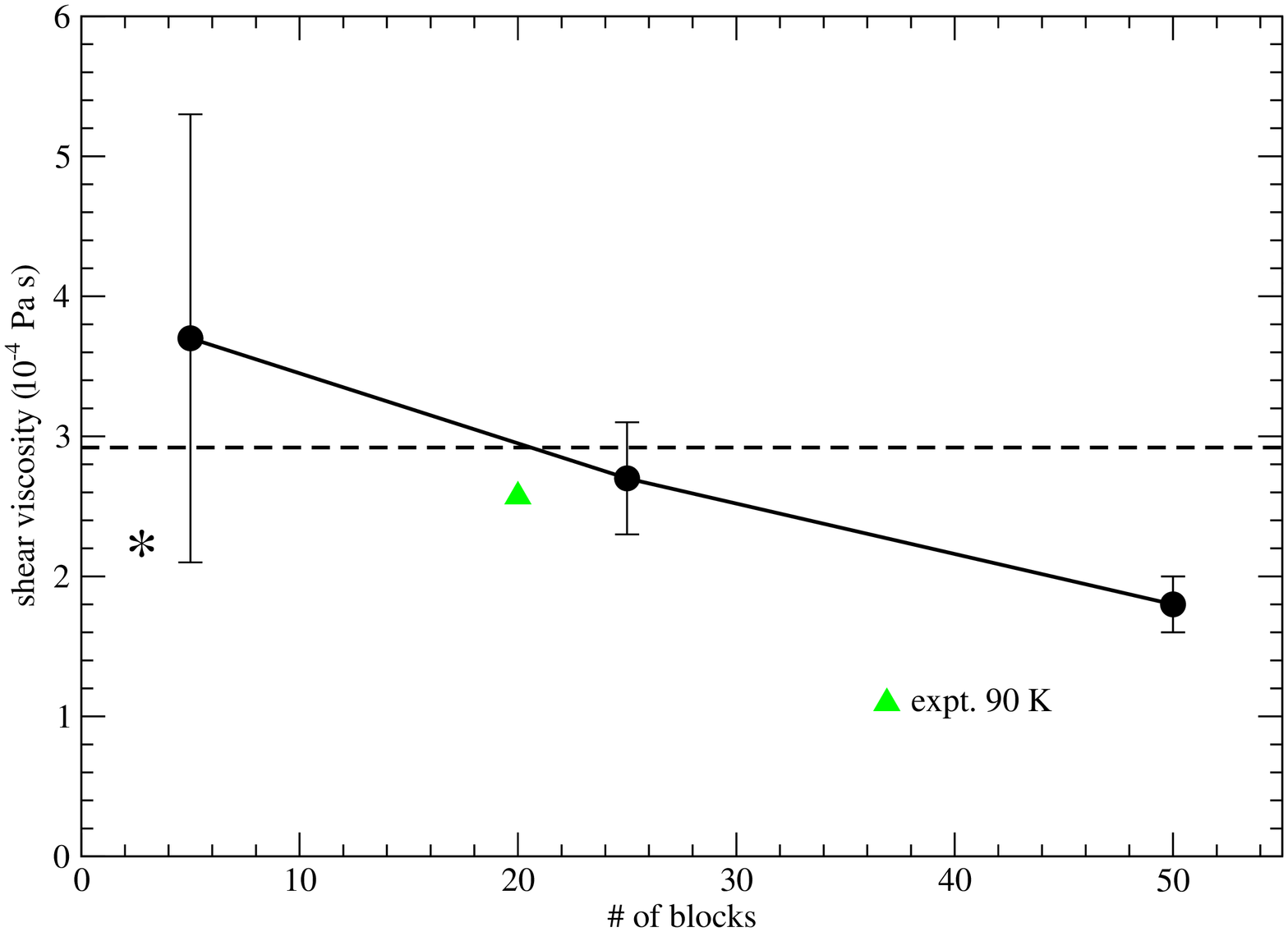}
}
\caption{Shear viscosity of liquid Ar evaluated by using 
different numbers of blocks (the smaller is the block number the larger 
is the number of configurations of each block) with the relative 
statistical errors.
The dashed horizontal line indicates the position of the first-pronounced 
maximum-plateau of the corresponding curve of Fig. 6. The asterisk denotes the value
obtained by the Stokes-Einstein formula (Eq.\ref{stokeseins}), while the
triangle indicates the experimental estimate at 90 K.}
\label{fig10}
\huge
\end{figure}
\pagestyle{empty}

\begin{figure}
{\vskip -1.0cm}
\centerline{
\includegraphics[width=17cm,angle=270]{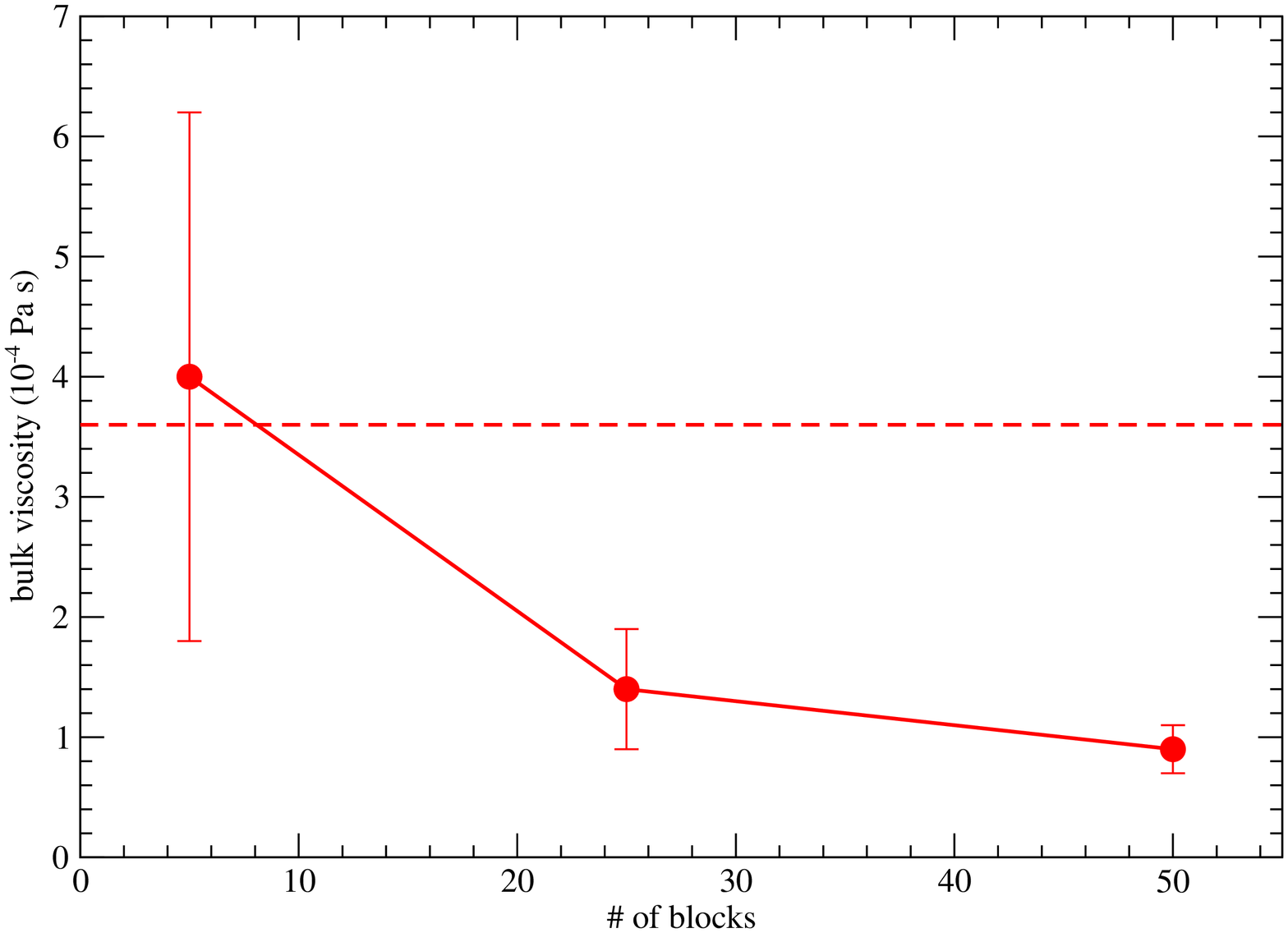}
}
\caption{Bulk viscosity of liquid Ar evaluated by using 
different numbers of blocks (the smaller is the block number the larger 
is the number of configurations of each block) with the relative 
statistical errors.
The dashed horizontal line indicates the position of the first-pronounced
maximum-plateau of the corresponding curve of Fig. 6.} 
\label{fig11}
\huge
\end{figure}

\pagestyle{empty}
\begin{figure}
{\vskip -1.0cm}
\centerline{
\includegraphics[width=17cm,angle=270]{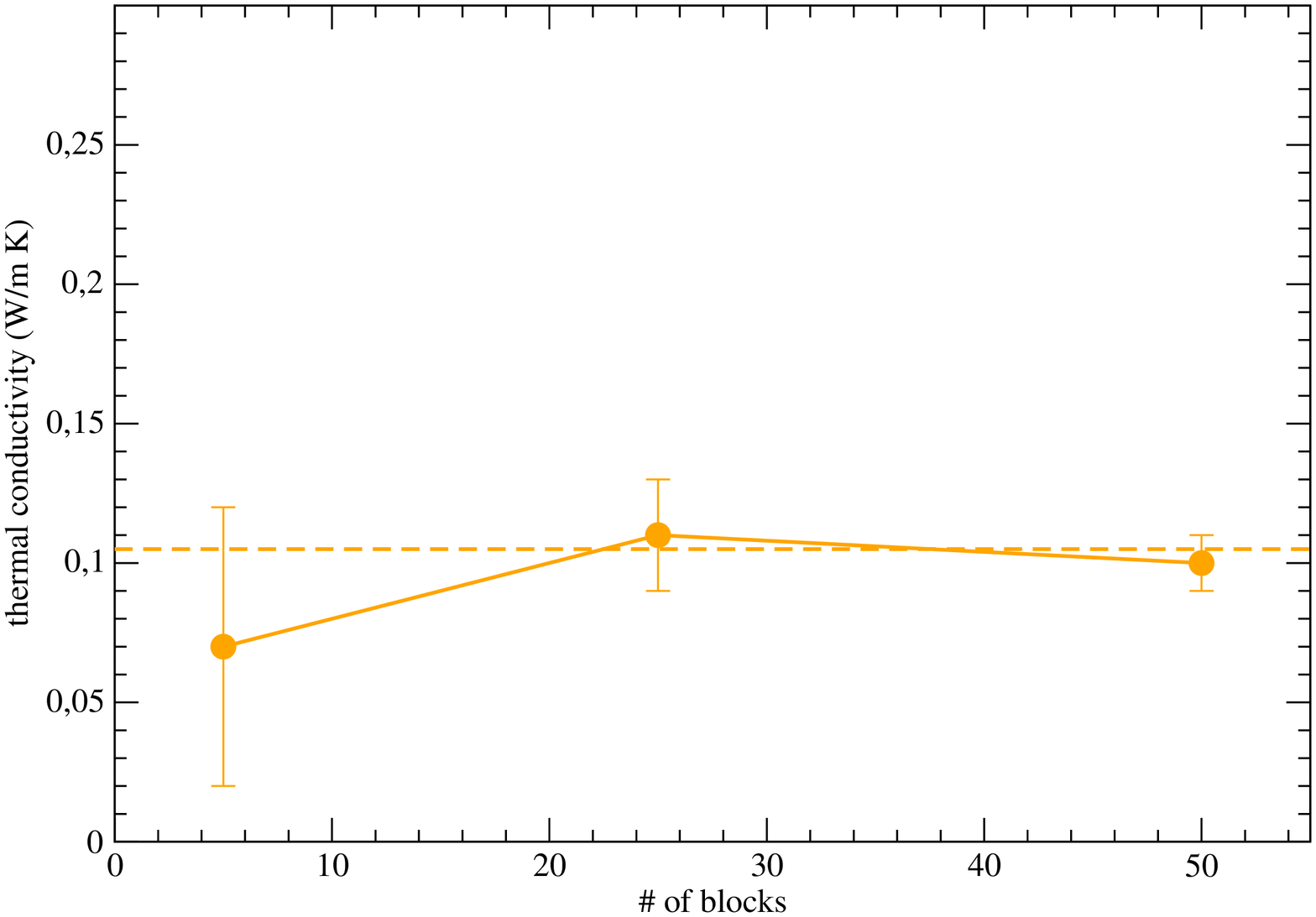}
}
\caption{Thermal conductivity of liquid Ar evaluated by using 
different numbers of blocks (the smaller is the block number the larger 
is the number of configurations of each block) with the relative 
statistical errors.
The dashed horizontal line indicates the position of the maximum-plateau
of the corresponding curve in Fig. 7.} 
\label{fig12}
\huge
\end{figure}

For liquid Ar the behavior is qualitatively similar (see Figs. 10, 11, and 12
for the shear viscosity, the bulk viscosity, and the thermal conductivity,
respectively). In this case both the maxima of the shear and bulk viscosities
are obtained considering 5 blocks, each equivalent to a simulation time of 
about 12.0 ps. 
Even in this case, taking statistical uncertainties into account, these maxima
are compatible with the plateau positions and, for the shear viscosity, 
also with the estimate from the Stokes-Einstein formula. 
The maximum of the thermal conductivity is instead reached with 25 blocks, each 
equivalent to a simulation time of about 2.4 ps, and its
value ($0.11 \pm 0.02$ W/m K) is again compatible with that
estimated considering the maximum-plateau position
and in good agreement with the literature reference value (0.12 W/m K) 
at 90 K\cite{Younglove} and that obtained by classical MD simulations 
based on the Lennard-Jones potential (0.119 W/m K).\cite{Hoheisel87}
Accurate first-principles calculations of the thermal conductivity of liquid Ar 
have been carried out, at higher temperatures (250 and 400 K), by  
Marcolongo {\it et al.}\cite{Marcolongo}, while the thermal conductivity of liquid 
water (adopting different DFT functionals) has been evaluated by Tisi {\it et al.}\cite{Tisi}
also using a deep neural network trained on DFT data. 

An interesting physical quantity is represented by the ratio between
bulk and shear viscosity, which can be related to the ratio of
observed to classical absorption coefficients in ultrasonic 
absorption experiments.\cite{Guo} In fact, under the condition that 
the heat conductivity contribution to the
ultrasonic absorption may be neglected,

\begin{equation}
{{\alpha} \over {\alpha_{class}}} = 3/4 {{\eta_B} \over {\eta_S}} + 1\;,
\label{absorption}
\end{equation} 

and water belongs to the group of the so-called ''associated liquids'',
characterized by a ratio from 1 to 3, where structural relaxation is 
dominant.

Classical MD simulations based on the SPC/E semiempirical potential
predict\cite{Guo} a ${\eta_B} \over {\eta_S}$ ratio of 2.4, leading
to a ${\alpha} \over {\alpha_{class}}$ ratio of 2.79, in reasonable agreement
with the experimental value of 3.0.\cite{Litovitz} 
Instead normal liquids, such as monatomic liquids (for instance
liquid Ar) are characterized by a ratio no greater than 1.2.
Although in general the ratio varies with temperature and pressure,
in liquid water it is found to remain constant within 20\% in the temperature 
range 0-90 C (273-363 K).\cite{Davis}
By taking statistical errors into account, our estimated value of the 
${\alpha} \over {\alpha_{class}}$ ratio ($2.8 \pm 0.6 $) is compatible 
with the available experimental data at ambient temperature (3.0).  
This is a remarkable result, considering that most of the reported classical 
MD simulations\cite{Guo} predict a bulk viscosity lower than the 
the experimental one, leading to an underestimated value of 
the ${\alpha} \over {\alpha_{class}}$ ratio. 

One should also point out that a proper comparison with experimental data 
requires a careful analysis taking into
account the pronounced temperature dependence of shear and bulk viscosity.
In fact, according to a common empirical model,\cite{Reynolds,Holmes} the 
viscosity strongly decreases with increasing temperature following 
an exponential decay.
By fitting experimental data\cite{Holmes} with an exponential function 
and taking statistical errors into account, our estimated values of 
the shear and bulk viscosity of liquid water
are compatible with experimental data in the temperature range of
323-344 K.
One should also consider that also the bulk-viscosity/shear-viscosity ratio 
for liquid water tends to decrease slightly with temperature,\cite{Holmes} 
suggesting an even better agreement between our estimated value and 
the experimental data.\cite{Holmes}
We remind that our simulations have been carried out at temperatures higher
than ambient temperature to guarantee that the systems is liquid-like.
By considering that our estimate (after finite-size correction) 
for the diffusion coefficient, $D=5.02\times 10^{-5}$ cm$^2$/s, 
corresponds to the experimental
value measured at about 336 K,\cite{diffcoeff} we can conclude that, 
our DFT simulations based
on the DFT-D2(BLYP) functional and performed at a nominal average temperature of
366 K, actually describe the basic dynamical properties of liquid water at about
330 K.
One should also mention that bulk-viscosity measurements are indirect and 
affected by considerable errors.\cite{Guo,Fernandez,Holmes,Dukhin,L-D,Xu}
In summary, we can conclude that our adopted BLYP-D2 functional is able to
describe reasonably well the density fluctuations of liquid water; 
the discrepancy with 
respect to experimental data at ambient conditions can be to a large 
extend explained in terms of the 
pronounced temperature dependence of both shear and bulk viscosity and the need 
to perform first-principles MD simulations at temperatures higher than 
ambient temperature, in line with the conclusions of other 
investigations.\cite{Malosso,Herrero}

As far as liquid Ar is concerned, our shear and bulk viscosities, computed by 
first-principles at a nominal average simulation temperature of 129 K, 
turn out to be
somehow overestimated with respect to the reference experimental values at 90 K,
although they are compatible with them if statistical errors are taken 
into account.
Moreover our bulk-viscosity/shear-viscosity ratio (close to unity) agrees well 
with the reference estimate, while interestingly this is not the case if 
a standard
Lennard-Jones empirical potential is adopted using classical MD simulations that
predict instead a very low value\cite{Guo,Hoheisel87} of the ratio 
(0.17-0.35 at high densities), thus showing that this popular potential cannot 
properly reproduce all the dynamical properties of liquid Ar and underlining
once again the superiority of first-principles approaches.     

We conclude our study by reporting some basic structural properties of our 
investigated systems. 
In particular, in Fig. 13, for liquid water we plot our computed O-O pair 
correlation function, $g_{OO}(r)$,
compared with that obtained experimental from X-ray diffraction measurements
at ambient conditions.\cite{Skinner,Skinnerbis,Daru}
The main features of the $g_{OO}(r)$ curves are reported in Table II.
As can be seen, there is a good agreement between the two curves; the
fact the oscillations of our computed curve are slightly reduced with 
respect to the experimental one can again be related to the higher 
effective temperature of our simulation.

\pagestyle{empty}
\begin{figure}
{\vskip -1.0cm}
\centerline{
\includegraphics[width=17cm,angle=270]{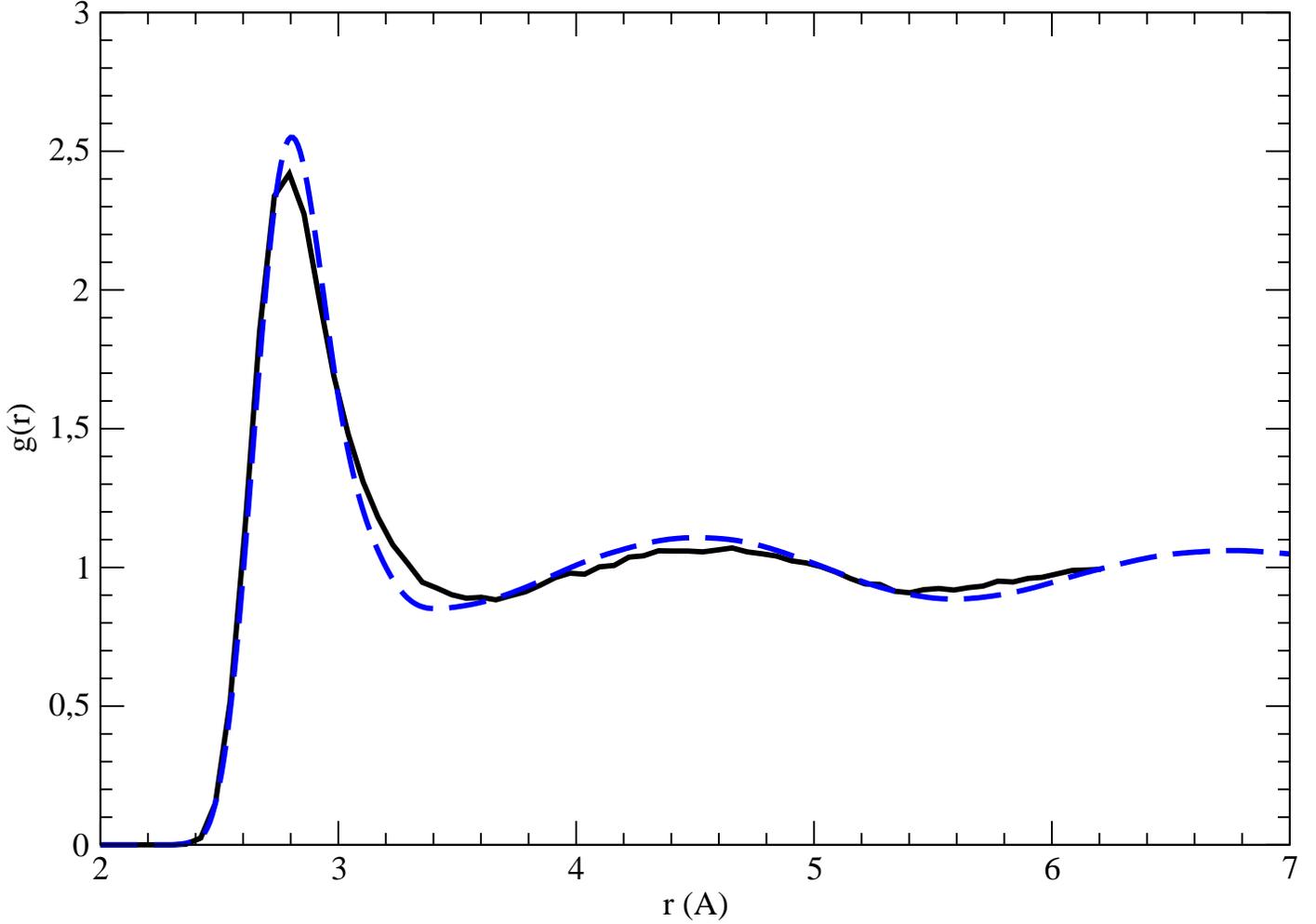}
}
\caption{O-O pair correlation function, $g_{OO}(r)$, compared 
with that obtained experimentally from X-ray diffraction measurements at
ambient conditions.\cite{Skinner,Skinnerbis,Daru}}
\label{fig13}
\huge
\end{figure}

\pagestyle{empty}
\begin{figure}
{\vskip -1.0cm}
\centerline{
\includegraphics[width=17cm,angle=270]{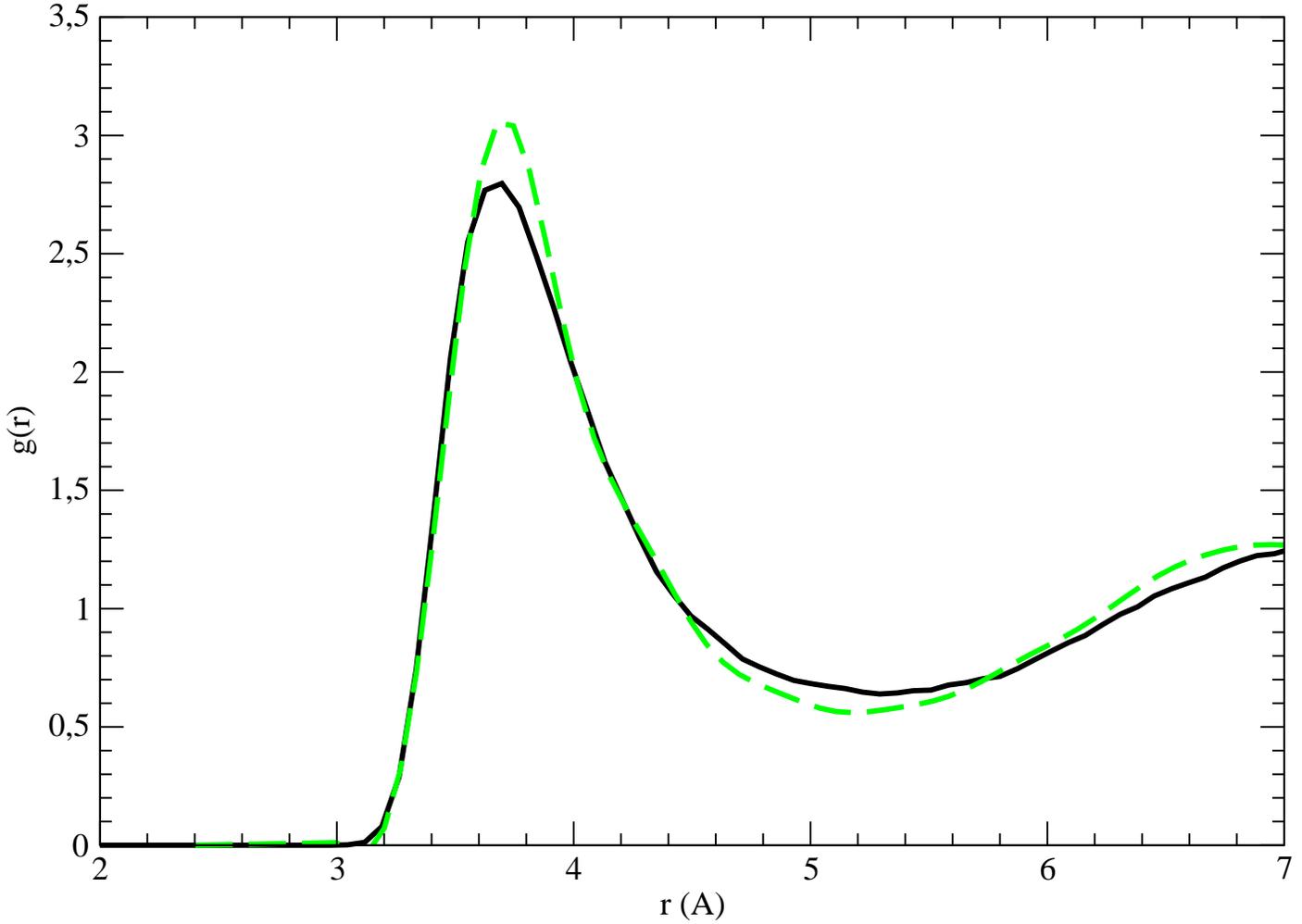}
}
\caption{Ar-Ar pair correlation function, $g(r)$, compared
with that obtained experimentally from neutron-scattering measurements 
at 85K.\cite{Yarnell}}
\label{fig14}
\huge
\end{figure}

\begin{table}
\vfill
\eject
\caption{Main features of the O-O pair correlation function, $g_{OO}(r)$,
of liquid water and of the Ar-Ar pair correlation function, $g(r)$ of 
liquid Ar compared with 
experimental reference data, obtained from X-ray diffraction measurements
at ambient conditions for liquid water and neutron-scattering measurements
for liquid Ar. 
$r_{max}$ and $r_{min}$ indicate the position of the first maximum (the main
peak) and the first minimum of $g_{OO}(r)$ and $g(r)$, respectively, and 
$g_{max}$ and $g_{min}$ the corresponding values of the $g_{OO}(r)$ 
and $g(r)$ functions.}
\begin{center}
\begin{tabular}{|c|c|c|c|c|}
\hline
system & $r_{max}$(\AA) & $g_{max}$& $r_{min}$(\AA) & $g_{min}$  \\ \tableline
\hline
 water (366 K)  & 2.79   & 2.42   & 3.66   & 0.88  \\
 water expt.$^a$ (293 K)      & 2.80(1)& 2.55(5)& 3.41(4)&0.85(2) \\
\hline
 Ar (129 K)          & 3.70   & 2.80   & 5.29   & 0.64  \\       
 Ar expt.$^b$ (85 K) & 3.68   & 3.05   & 5.18   & 0.56  \\
\hline
\end{tabular}                                                
\tablenotetext[1]{ref.\onlinecite{Skinner,Skinnerbis,Daru}.}
\tablenotetext[2]{ref.\onlinecite{Yarnell}.}
\end{center}
\label{table2}                                  
\end{table}
\eject

In Fig. 14, for liquid Ar our computed Ar-Ar pair 
correlation function, $g(r)$, is
compared with that obtained experimentally from neutron-scattering measurements
at 85 K,\cite{Yarnell} while again
the main features of the $g(r)$ curves are reported in Table II.
Even in this case there is a reasonable agreement between the simulation and
experimental curve, by considering that simulations for liquid Ar have been
performed at significantly higher temperature (129 K) than experiments (85 K)
(note that the experimental melting and boiling points of Ar are at 
84 and 87 K, respectively).
After applying the same finite-size correction adopted above for
liquid water, our estimated diffusion coefficient for liquid Ar, 
$D=3.82\times 10^{-5}$ cm$^2$/s, evaluated at a nominal simulation temperature
of 129 K is significantly higher than the reference value 
($1.6\times 10^{-5}$ cm$^2$/s) reported at 84 K.\cite{Hansen}
Again this discrepancy can be explained in terms of the higher 
temperature of the liquid Ar simulation.

\section{Conclusions}
Shear and bulk viscosity of liquid water and Argon have been evaluated,
together with other structural and dynamical properties, from first
principles by adopting a vdW-corrected DFT approach, by performing
Molecular Dynamics simulations in the NVE ensemble and using
the Kubo-Greenwood equilibrium approach.
For liquid Argon the thermal conductivity has been also calculated.
Concerning liquid water, to our knowledge this is the first
estimate of the bulk viscosity and of the 
shear-viscosity/bulk-viscosity ratio from first principles.
By analyzing our results and comparing then with reference experimental data,
we can conclude that our first-principles simulations,
performed at a nominal average temperature of
366 K to guarantee that the systems is liquid-like, actually describe 
well the basic dynamical properties of liquid water at about 330 K.
In comparison with liquid water, the normal, monatomic liquid Ar 
is characterized by a much smaller bulk-viscosity/shear-viscosity ratio 
(close to unity) and this feature is well reproduced by our first-principles 
approach which predicts a value of the ratio in better agreement 
with experimental reference data than that obtained using the 
empirical Lennard-Jones potential.  
The computed thermal conductivity of liquid Argon is also in good agreement with
the experimental value. 

\section{Acknowledgements}
We acknowledge S. Baroni for useful disucssions and 
funding from Fondazione Cariparo, Progetti di Eccellenza 2017,
relative to the project: ''Engineering van der Waals
Interactions: Innovative paradigm for the control of Nanoscale
Phenomena''.

\section{Data availability}
The data that support the findings of this study are available from the 
corresponding author upon reasonable request.

\noindent
\narrowtext

\vfill
\eject

\end{document}